\documentclass{article}
\usepackage{amssymb}
\usepackage{graphicx}
\usepackage{amsmath}

\input{tcilatex}

\begin{document}

\title{Constants and Variations: \\
From Alpha to Omega}
\author{John D. Barrow \\
DAMTP\\
Centre for Mathematical Sciences\\
Cambridge University\\
Cambridge CB3 0WA\\
UK}
\maketitle

\begin{abstract}
We review some of the history and properties of theories for the variation
of the gravitation and fine structure 'constants'. We highlight some general
features of the cosmological models that exist in these theories with
reference to recent quasar data that is consistent with time-variation in
alpha since a redshift of 3.5. The behaviour of a simple class of
varying-alpha cosmologies is outlined in the light of all the observational
constraints. We discuss the key role played by non-zero vacuum energy and
curvature in turning off the variation of constants in these theories and
the issue of comparing extra-galactic and local observational data. We also
show why black hole thermodynamics does not enable us to distinguish between
time variations of different constants.
\end{abstract}

\section{ Introduction}

\bigskip\ \ 

There are a number of reasons why the possibility of varying constants
should be taken seriously \cite{con}. First, we know that the best
candidates for unification of the forces of nature in a quantum
gravitational environment only seem to exist in finite form if there are
many more dimensions of space than the three that we are familiar with. This
means that the true constants of nature are defined in higher dimensions and
the three-dimensional shadows\ we observe are not fundamental and do not
need to be constant. Any slow change in the scale of the extra dimensions
would be revealed by measurable changes in our three-dimensional
'constants'. Second, we appreciate that some apparent constant might be
determined partially or completely by some spontaneous symmetry-breaking
processes in the very early universe. This introduces an irreducible random
element into the values of those constants. They may be different in
different parts of the universe. The most dramatic manifestation of this
process is provided by the chaotic and eternal inflationary universe
scenarios. Third, any outcome of a theory of quantum gravity will be
intrinsically probabilistic. It is often imagined that the probability
distributions for observables will be very sharply peaked but this may not
be the case for all possibilities. Thus, the value of $G$ or $\dot{G}$ might
be predicted to be spatially varying random variables. Fourth, the
non-uniqueness of the vacuum state for the universe would allow other deals
of the constants to have occurred in different places. At present we have no
idea why any of the constants of Nature take the numerical values they do.
Fifth, the observational limits on possible variations are often very weak
(although they can be made to sound strong by judicious parametrisations).
For example, the cosmological limits on varying $G$ tell us only that $\dot{G%
}/G\leq 10^{-2}H_{0}$, where $H_{0}$ is the present Hubble rate. However,
the last reason to consider varying constants is currently the most
compelling. For the first time there is a body of detailed astronomical
evidence for the time variation of a traditional constant. The observational
programme of Webb et al \cite{webb1,webb2} has completed detailed analyses
of three separate quasar absorption line data sets taken at Keck and finds
persistent evidence consistent with the fine structure constant, $\alpha $,
having been \textit{smaller} in the past, at $z=1-3.5.$ The shift in the
value of $\alpha $ for all the data sets is given provisionally by $\Delta
\alpha /\alpha =(-0.66\pm $ $0.11)\times 10^{-5}.$ This result is currently
the subject of detailed analysis and reanalysis by the observers in order to
search for possible systematic biases in the astrophysical environment or in
the laboratory determinations of the spectral lines.

The first investigations of time-varying constants were those made by Lord
Kelvin and others interested in possible time-variation of the speed of
light at the end of the nineteenth century. In 1935 Milne devised a theory
of gravity, of a form that we would now term 'bimetric', in which there were
two times -- one ($t$) for atomic phenomena, one ($\tau $) for gravitational
phenomena -- linked by $\tau =\log (t/t_{0})$. Milne \cite{mil} required
that the 'mass of the universe' (what we would now call the mass inside the
particle horizon $M\ \approx c^{3}G^{-1}t$) be constant. This required $%
G\varpropto t.$ Interestingly, in 1937 the biologist J.B.S. Haldane took a
strong interest in this theory and wrote several papers \cite{hal} exploring
its consequences for the evolution of life. The argued that biochemical
activation energies might appear constant on the $t$ timescale yet increase
on the $\tau $ timescale, giving rise to a non-uniformity in the
evolutionary process. Also at this time there was widespread familiarity
with the mysterious 'large numbers' $O(10^{40})$ and $O(10^{80})$ through
the work of Eddington (although they had first been noticed by Weyl \cite%
{weyl} -- see ref. \cite{bt} and \cite{con} for the history). These two
ingredients were merged by Dirac in 1937 in a famous development (supposedly
written on his honeymoon) that proposed that these large numbers $(10^{40})$
were actually equal, up to small dimensionless factors. Thus, if we form $%
N\sim c^{3}t/Gm_{n}\sim 10^{80}$, the number of nucleons in the visible
universe, \ and equate it to the square of $N_{1}\sim e^{2}/Gm_{n}^{2}\sim
10^{40},$ the ratio of the electrostatic and gravitational forces between
two protons then we are led to conclude that one of the constants, $%
e,G,c,h,m_{n}$ must vary with time. Dirac \cite{Dir} chose $G\varpropto
t^{-1}$ to carry the time variation. Unfortunately, this hypothesis did not
survive very long. Edward Teller \cite{tell} pointed out that such a steep
increase in $G$ to the past led to huge increases in the Earth's surface
temperature in the past. The luminosity of the sun varies as $L\varpropto
G^{7}$ and the radius of the Earth's orbit as $R\varpropto G^{-1}$ so the
Earth's surface temperature $T_{\oplus }$varies as $(L/R^{2})^{1/4}%
\varpropto G^{9/4}\varpropto t^{-9/4}$ and would exceed the boiling point of
water in the pre-Cambrian era. Life would be eliminated. Gamow subsequently
suggested that the time variation needed to reconcile the large number
coincidences be carried by $e$ rather than $G$, but again this strong
variation was soon shown to be in conflict with geophysical and radioactive
decay data. This chapter was brought to an end by Dicke \cite{dicke} who
pointed out that the $N\sim N_{1}^{2}$ large number coincidence was just the
statement that $t$, the present age of the universe when our observations
are being made, is of order the main sequence stellar lifetime, $t_{ms}\sim
(Gm_{n}^{2}/hc)^{-1}h/m_{n}c^{2}\sim 10^{10}yrs$, and therefore inevitable
for observers made from elements heavier than hydrogen and helium. Dirac
never accepted this anthropic explanation for the large number coincidences
but curiously can be found making exactly the same type of anthropic
argument to defend his own varying $G$ theory by highly improbable arguments
(that the Sun accretes material periodically during its orbit of the galaxy
and this extra material cancels out the effects of overheating in the past)
in correspondence with Gamow in 1967 (see \cite{con} for fuller details).

Dirac's proposal acted as a stimulus to theorists, like Jordan, Brans and
Dicke \cite{bd}, to develop rigorous theories which included the time
variation of $G$ self-consistently by modelling it as arising from the
space-time variation of some scalar field $\phi (\mathbf{x},t)$ whose motion
both conserved energy and momentum and created its own gravitational field
variations. In this respect the geometric structure of Einstein's equations
provides a highly constrained environment to introduce variations of
'constants'. Whereas in Newtonian gravity we are at liberty to introduce a
time-varying $G(t)$ into the law of gravity by

\begin{equation}
F=-\frac{G(t)Mm}{r^{2}}  \label{newt1}
\end{equation}

This creates a non-conservative dynamical system but can be solved fairly
straightforwardly \cite{jbnewt}. However, this strategy of simply 'writing
in' the variation of $G$ by merely replacing $G$ by $G(t)$ in the equations
that hold when $G$ is a constant fails in general relativity. If we were to
imagine the Einstein equations would generalise to ($G_{ab}$ is the Einstein
tensor)

\begin{equation}
G_{ab}=\frac{8\pi G(t)}{c^{4}}T_{ab}  \label{newt}
\end{equation}%
then taking a covariant divergence and using $\nabla ^{a}G_{ab}=0,$ together
with energy-momentum conservation ($\nabla ^{a}T_{ab}=0)$ requires that $%
\nabla G\equiv 0$ and no variations are possible in eq. (\ref{newt}).
Brans-Dicke theory is a familiar example of how the addition of an extra
piece to $T_{ab\text{ \ }}$together with the dynamics of a $G(\phi )$ fields
makes a varying $G$ theory possible. Despite the simplicity of this lesson
in the context of a varying $G$ theory the lesson was not taken on board
when considering the variations of other non-gravitational constants and the
literature is full of limits on their possible variation which have been
derived by considering a theory in which the time-variation is just written
into the equations which hold when the constant does not vary. Recently, the
interest in the possibility that $\alpha $ varies in time has led to the
first extensive exploration of simple self-consistent theories in which $a$
variations occur through the variation of some scalar field.

\section{Brans-Dicke Theories}

\subsection{\protect\bigskip Equations and solutions}

\emph{\ }Consider the paradigmatic case of Brans-Dicke (BD) theory \cite{bd}
to fix theoretical ideas about varying $G$. The form of the general
solutions to the Friedmann metric in BD theories are fully understood \cite%
{bar}, \cite{fink}. There are three essential field equations for the
evolution of BD scalar field $\phi (t)$ and the expansion scale factor $a(t)$
in a BD universe

\begin{eqnarray}
3\frac{\dot{a}^{2}}{a^{2}} &=&\frac{8\pi \rho }{\phi }-3\frac{\dot{a}^{\ }%
\dot{\phi}}{a\ \phi }+\frac{\omega _{BD}}{2}\frac{\dot{\phi}^{2}}{\phi ^{2}}-%
\frac{k}{a^{2}}  \label{bd1} \\
\ddot{\phi}+3\frac{\dot{a}}{a}\dot{\phi} &=&\frac{8\pi }{3+2\omega _{BD}}%
(\rho -3p)  \label{bd2} \\
\dot{\rho}+3\frac{\dot{a}}{a}(\rho +p) &=&0  \label{bd3}
\end{eqnarray}
Here, $\omega _{BD}$ is the BD constant parameter and the theory reduces to
general relativity in the limit $\omega _{BD}\rightarrow \infty $ and $\phi
=G^{-1}\rightarrow $ constant. A general feature of the BD field equations
is that any solution of general relativity for which the energy momentum
tensor of matter has vanishing trace (eg vacuum, black body radiation,
Yang-Mills, or magnetic field) is a particular ($\phi =$ constant) solution
of BD theory.

The general solutions begin at high density dominated by the BD scalar field 
$\phi \sim G^{-1}$ and approximated are well approximated by the spatially
flat vacuum ($\rho =p=0$) solutions:

\begin{eqnarray}
a(t) &=&t^{1/(\lambda +1)}  \label{sol1} \\
\phi (t) &=&\phi _{0}t^{\lambda /(\lambda +1)}  \label{sol2} \\
\lambda &=&\frac{1+\sqrt{1+2\omega _{BD}/3}}{\omega _{BD}}  \label{sol3}
\end{eqnarray}%
This vacuum solution is the $t\rightarrow 0$ attractor for the perfect-fluid
solutions. The general perfect-fluid solutions with equation of state $%
p=\Gamma \rho $ and $k=0$ can all be found. At early times they approach the
vacuum solutions but at late time they approach particular power-law exact
solutions \cite{nar}:

\begin{equation}
a(t)=t^{[2+2\omega _{BD}(1-\Gamma )]/[4+3\omega _{BD}(1-\Gamma ^{2})]}
\label{bds1}
\end{equation}

\begin{equation}
\phi (t)=\phi _{0}t^{[2(1-3\Gamma )/[4+3\omega (1-\Gamma ^{2})]}
\label{bds2}
\end{equation}

At late times they approach particular exact power-law solutions for $a(t)$
and $\phi (t)$ and the evolution is 'Machian' in the sense that the
cosmological evolution is driven by the matter content rather than by the
kinetic energy of the free $\phi $ field.

In the radiation era this particular solution is the standard general
relativity solution:

\begin{equation}
a(t)=t^{1/2};\hspace{1in}\phi ^{-1}\propto G=constant  \label{r}
\end{equation}
For $p=0$ the solutions have the form

\begin{equation}
a(t)=t^{(2-n)/3};\hspace{1in}\phi ^{-1}\varpropto G\propto t^{-n},
\label{dust}
\end{equation}%
which continues until the curvature term takes over the expansion. Here, $n$
is related to the constant Brans-Dicke $\omega _{BD}$ parameter by

\begin{equation}
n\equiv \frac{2}{4+3\omega _{BD}}  \label{n}
\end{equation}
and the usual general relativistic Einstein de Sitter universe is obtained
as $\ \omega _{BD}\rightarrow \infty $ and $n\rightarrow 0.$

In a curvature-dominated era of expansion, as $t\rightarrow \infty $ the
solutions for $\Gamma >-1/3$ approach the general relativity Milne vacuum
solution with

\begin{eqnarray}
a(t) &=&t \\
\phi &\varpropto &G^{-1}=\text{constant}
\end{eqnarray}

Notice how the curvature domination turns off the variation of $G$. All
previous studies of varying $G$ in cosmology have focussed on the $k=0$
models and have not noticed the important role that would be played by
negative curvature. The existence of negative curvature can produce little
residue of $G$ variation in the universe today. It also highlights the
usefulness of having constraints from different cosmic times and redshifts.

The other important lesson to learn from the cosmological limits on varying $%
G$ is that care must be taken when using local limits on $\omega _{BD}$, say
from light-bending by the Sun or the other solar system tests of general
relativity, and then assuming that they can be used in cosmological models.
In reality the evolution of the universe is inhomogeneous and there are very
large variations in density between the solar system and the extragalactic
universe. If we had a perfect numerical simulation of cosmology with a
varying $G$ we would be able to determine the contours of $G$ and $\dot{G}$
with position in the universe. Until we have more information of that sort
from models it is unwise to assume that the rate of change of $G$ in the
solar system will be the same as it is on cosmological scales.

An interesting particular example of this problem is given by the power-law
solutions above for the case with $\Gamma =-1.$ This is equivalent to the
universe being dominated by a vacuum energy and leads to power-law
accelerated expansion in BD theory with

\begin{eqnarray}
a(t) &=&t^{\frac{1}{2}+\omega } \\
&&G\varpropto \phi ^{-1}\varpropto t^{-2}
\end{eqnarray}%
Thus it appears that if our universe were to be expanding today (as
observations of the recession of type I supernovae indicate \cite{super})
then $G$ must be falling very rapidly (even faster than Dirac predicted)
locally. Clearly this is impossible observationally \cite{will}. The flaw in
the argument is that the $p=-\rho $ stress does not dominate the dynamics of
the solar system and we must not apply the cosmological solution for the
variation of $G$ in the solar system anymore than we should apply the
cosmological solution for the variation of $\rho $ to the solar system.

It would be very interesting to find realistic solutions (exact,
approximate, or numerical) for inhomogeneous cosmological models with $\phi (%
\mathbf{x},t)$ in order to obtain some perspective on the likely variation
in the change in $G$ from solar system to galactic and extragalactic scales.
At present potentially the strongest cosmological limit on time-varying $G$
is stronger than solar system tests and comes, somewhat surprisingly, from
the power spectrum of galaxy clustering. The effect of varying $G$ is to
shift the cosmic epoch of equality between the matter and radiation
densities which determines the location of the peak of the clustering power
spectrum \cite{lidd}.

\section{A Simple Varying-Alpha Theory}

\bigskip

We are going to consider some of the cosmological consequences of a simple
theory with time varying $\alpha .$ Such a theory was first formulated by
Bekenstein \cite{bek} as a generalisation of Maxwell's equations but
ignoring the consequences for the gravitational field equations. Recently,
Magueijo, Sandvik and myself have completed this theory \ \cite{bsm1, bsm2,
bsm3, bsm4} to include the coupling to the gravitational sector and analysed
its general cosmological consequences. This theory considers only a
variation of the electromagnetic coupling and so far ignores any unification
with the strong and electroweak interactions. We shall not discuss
simultaneous variation of the electromagnetic and gravitational constants
although that analysis can be done and is presented elsewhere (during the
dust era of a flat Friedmann universe with varying $\alpha (t)$ and $G(t),$
their time-evolution approaches an attractor in which the product $\alpha G$
is a constant and $\alpha \propto G^{-1}\propto t^{n},$ where $n$ is given
by eq. (\ref{n}).)

Our aim in studying this theory is to build up understanding of the effects
of the expansion on varying $\alpha $ and to identify features that might
carry over into more general theories in which all the unified interactions
vary \cite{banks, guts, marc}. The constraint imposed on varying $\alpha $
by the need to bring about unification at high energy is likely to be
significant but the complexities of analysing the simultaneous variation of
all the constants involved in the supersymmetric version of the standard
model are considerable. At the most basic level we recognise that any time
variation in the fine structure could be carried by either or both of the
electromagnetic or weak couplings above the electroweak scale.

The idea that the charge on the electron, or the fine structure constant,
might vary in cosmological time was proposed in 1948 by Teller, \cite{tell},
who suggested that $\alpha \propto (\ln t)^{-1}$ was implied by Dirac's
proposal that $G\propto t^{-1}$ and the numerical coincidence that $\alpha
^{-1}\sim \ln (hc/Gm_{pr}^{2})$, where $m_{pr\text{ }}$is the proton mass.
Later, in 1967, Gamow \cite{gam} suggested $\alpha \propto t$ as an
alternative to Dirac's time-variation of the gravitation constant, $G$, as a
solution of the large numbers coincidences problem and in 1963 Stanyukovich
had also considered varying $\alpha $, \cite{stan}, in this context.
However, this power-law variation in the recent geological past was soon
ruled out by other evidence \cite{dyson}.

There are a number of possible theories allowing for the variation of the
fine structure constant, $\alpha $. In the simplest cases one takes $c$ and $%
\hbar $ to be constants and attributes variations in $\alpha $ to changes in 
$e$ or the permittivity of free space (see \cite{am} for a discussion of the
meaning of this choice). This is done by letting $e$ take on the value of a
real scalar field which varies in space and time (for more complicated
cases, resorting to complex fields undergoing spontaneous symmetry breaking,
see the case of fast tracks discussed in \cite{covvsl}). Thus $%
e_0\rightarrow e=e_0\epsilon (x^\mu ),$ where $\epsilon $ is a dimensionless
scalar field and $e_0$ is a constant denoting the present value of $e$. This
operation implies that some well established assumptions, like charge
conservation, must give way \cite{land}. Nevertheless, the principles of
local gauge invariance and causality are maintained, as is the scale
invariance of the $\epsilon $ field (under a suitable choice of dynamics).
In addition there is no conflict with local Lorentz invariance or covariance.

With this set up in mind, the dynamics of our theory is then constructed as
follows. Since $e$ is the electromagnetic coupling, the $\epsilon $ field
couples to the gauge field as $\epsilon A_{\mu }$ in the Lagrangian and the
gauge transformation which leaves the action invariant is $\epsilon A_{\mu
}\rightarrow \epsilon A_{\mu }+\chi _{,\mu },$ rather than the usual $A_{\mu
}\rightarrow A_{\mu }+\chi _{,\mu }.$ The gauge-invariant electromagnetic
field tensor is therefore 
\begin{equation}
F_{\mu \nu }=\frac{1}{\epsilon }\left( (\epsilon A_{\nu })_{,\mu }-(\epsilon
A_{\mu })_{,\nu }\right) ,
\end{equation}
which reduces to the usual form when $\epsilon $ is constant. The
electromagnetic part of the action is still 
\begin{equation}
S_{em}=-\int d^{4}x\sqrt{-g}F^{\mu \nu }F_{\mu \nu }.
\end{equation}
and the dynamics of the $\epsilon $ field are controlled by the kinetic term 
\begin{equation}
S_{\epsilon }=-\frac{1}{2}\frac{\hslash }{l^{2}}\int d^{4}x\sqrt{-g}\frac{%
\epsilon _{,\mu }\epsilon ^{,\mu }}{\epsilon ^{2}},
\end{equation}
as in dilaton theories. Here, $l$ is the characteristic length scale of the
theory, introduced for dimensional reasons. This constant length scale gives
the scale down to which the electric field around a point charge is
accurately Coulombic. The corresponding energy scale, $\hbar c/l,$ has to
lie between a few tens of $MeV$ and Planck scale, $\sim 10^{19}GeV$ to avoid
conflict with experiment.

Our generalisation of the scalar theory proposed by Bekenstein \cite{bek}
described in ref. \ \cite{bsm1, bsm2, bsm3, bsm4} includes the gravitational
effects of $\psi $ and gives the field equations: 
\begin{equation}
G_{\mu \nu }=8\pi G\left( T_{\mu \nu }^{matter}+T_{\mu \nu }^{\psi }+T_{\mu
\nu }^{em}e^{-2\psi }\right) .  \label{ein}
\end{equation}%
The stress tensor of the $\psi $ field is derived from the lagrangian $%
\mathcal{L}_{\psi }=-{\frac{\omega }{2}}\partial _{\mu }\psi \partial ^{\mu
}\psi $ and the $\psi $ field obeys the equation of motion 
\begin{equation}
\square \psi =\frac{2}{\omega }e^{-2\psi }\mathcal{L}_{em}  \label{boxpsi}
\end{equation}%
where we have defined the coupling constant $\omega =(c)/l^{2}$. This
constant is of order $\sim 1$ if, as in \cite{sbm}, the energy scale is
similar to Planck scale. It is clear that $\mathcal{L}_{em}$ vanishes for a
sea of pure radiation since then $\mathcal{L}_{em}=(E^{2}-B^{2})/2=0$. We
therefore expect the variation in $\alpha $ to be driven by electrostatic
and magnetostatic energy-components rather than electromagnetic radiation.

In order to make quantitative predictions we need to know how much of the
non-relativistic matter contributes to the RHS of Eqn.~(\ref{boxpsi}). This
is parametrised by $\zeta \equiv \mathcal{L}_{em}/\rho $, where $\rho $ is
the energy density, and for baryonic matter $\mathcal{L}_{em}=E^{2}/2$. For
protons and neutrons $\zeta _{p}$ and $\zeta _{n}$ can be \textit{estimated}
from the electromagnetic corrections to the nucleon mass, $0.63$ MeV and $%
-0.13$ MeV, respectively \cite{zal}. This correction contains the $E^{2}/2$
contribution (always positive), but also terms of the form $j_{\mu }a^{\mu }$
(where $j_{\mu }$ is the quarks' current) and so cannot be used directly.
Hence we take a guiding value $\zeta _{p}\approx \zeta _{n}\sim 10^{-4}$.
Furthermore the cosmological value of $\zeta $ (denoted $\zeta _{m}$) has to
be weighted by the fraction of matter that is non-baryonic, a point ignored
in the literature \cite{bek}. Hence, $\zeta _{m}$ depends strongly on the
nature of the dark matter and can take both positive and negative values
depending on which of Coulomb-energy or magnetostatic energy dominates the
dark matter of the Universe. It could be that $\zeta _{CDM}\approx -1$
(superconducting cosmic strings, for which $\mathcal{L}_{em}\approx
-B^{2}/2) $, or $\zeta _{CDM}\ll 1$ (neutrinos). BBN predicts an approximate
value for the baryon density of $\Omega _{B}\approx 0.03$ with a Hubble
parameter of $h_{0}\approx 0.6$ , implying $\Omega _{CDM}\approx 0.3$. Thus
depending on the nature of the dark matter $\zeta _{m}$ can be virtually
anything between $-1$ and $+1$. The uncertainties in the underlying quark
physics and especially the constituents of the dark matter make it difficult
to impose more certain bounds on $\zeta _{m}$.

We should not confuse this theory with other similar variations.
Bekenstein's theory does not take into account the stress energy tensor of
the dielectric field in Einstein's equations, and their application to
cosmology. Dilaton theories predict a global coupling between the scalar and
all other matter fields. As a result they predict variations in other
constants of nature, and also a different dynamics to all the matter coupled
to electromagnetism. An interesting application of our approach has also
recently been made to braneworld cosmology in \cite{youm}.

\subsection{The cosmological equations}

Assuming a homogeneous and isotropic Friedmann metric with expansion scale
factor $a(t)$ and curvature parameter $k$ in eqn. (\ref{ein}), we obtain the
field equations ($c\equiv 1$) 
\begin{eqnarray}
\left( \frac{\dot{a}}{a}\right) ^{2} &=&\frac{8\pi G}{3}\left( \rho
_{m}\left( 1+\zeta _{m}\exp {[-2\psi ]}\right) +\rho _{r}\exp {[-2\psi ]}+%
\frac{\omega }{2}\dot{\psi}^{2}\right)  \notag \\
&&-\frac{k}{a^{2}}+\frac{\Lambda }{3},  \label{fried1}
\end{eqnarray}
where $\Lambda $ is the cosmological constant. For the scalar field we have
the propagation equation, 
\begin{equation}
\ddot{\psi}+3H\dot{\psi}=-\frac{2}{\omega }\exp {[-2\psi ]}\zeta _{m}\rho
_{m},  \label{psidot}
\end{equation}
where $H\equiv \dot{a}/a$ is the Hubble expansion rate$.$ We can rewrite
this more simply as

\begin{equation}
(\dot{\psi}a^{3}\dot{)}=N\exp [-2\psi ]  \label{psidot2}
\end{equation}%
where $N$ is a positive constant defined by

\begin{equation}
N=-\frac{2\zeta _{m}\rho _{m}a^{3}}{\omega }  \label{N}
\end{equation}

Note that the sign of the evolution of $\psi $ is dependent on the sign of $%
\zeta _{m}$. Since the observational data is consistent with a \emph{smaller}
value of $\alpha $ in the past, we will in this paper confine our study to 
\emph{negative} values of $\zeta _{m}$, in line with our recent discussion
in Refs. \cite{bsm1, bsm2, bsm3, bsm4}. The conservation equations for the
non-interacting radiation and matter densities are 
\begin{eqnarray}
\dot{\rho _{m}}+3H\rho _{m} &=&0 \\
\dot{\rho _{r}}+4H\rho _{r} &=&2\dot{\psi}\rho _{r}.  \label{conservation}
\end{eqnarray}%
and so $\rho _{m}\propto a^{-3}$ and $\rho _{r}$ $e^{-2\psi }\propto a^{-4},$
respectively. If additional non-interacting perfect fluids satisfying
equation of state $p=(\gamma -1)\rho $ are added to the universe then they
contribute density terms $\rho \propto a^{-3\gamma }$ to the RHS of eq.(\ref%
{fried1}) as usual. This theory enables the cosmological consequences of
varying $e$, to be analysed self-consistently rather than by changing the
constant value of $e$ in the standard theory to another constant value, as
in the original proposals made in response to the large numbers coincidences.

We have been unable to solve these equations in general except for a few
special cases \cite{bmota}. However, as with the Friedmann equation of
general relativity, it is possible to determine the overall pattern of
cosmological evolution in the presence of matter, radiation, curvature, and
positive cosmological constant by matched approximations. We shall consider
the form of the solutions to these equations when the universe is
successively dominated by the kinetic energy of the scalar field $\psi $,
pressure-free matter, radiation, negative spatial curvature, and positive
cosmological constant$.$ Our analytic expressions are checked by numerical
solutions of (\ref{fried1}) and (\ref{psidot}).

\subsection{Observational implications}

\bigskip

There are a number of conclusions that can be drawn from the study of the
simple BSBM models with $\zeta _{m}<0$. These models give a good fit to the
varying $\alpha $ implied by the QSO data of refs. \cite{webb1,webb2}. There
is just a single parameter to fit and this is given by the choice

\begin{equation}
-\frac{\zeta _{m}}{\omega }=(2\pm 1)\times 10^{-4}  \label{om}
\end{equation}

The simple solutions predict a slow (logarithmic) time increase during the
dust era of $k=0$ Friedmann universes. The cosmological constant turns off
the time-variation of $\alpha $ at the redshift when the universe begins to
accelerate ($z\sim 0.7$) and so there is no conflict between the $\alpha $
variation seen in quasars at $z\sim 1-3.5$ and the limits on possible
variation of $\alpha $ deduced from the operation of the Oklo natural
reactor \cite{oklo} (even assuming that the cosmological variation applies
unchanged to the terrestrial environment). The reactor operated 1.8 billion
years ago at a redshift of only $z\sim 0.1$ when no significant variations
were occurring in $\alpha $. The slow logarithmic increase in $\alpha $ also
means that we would not expect to have seen any effect yet in the anisotropy
of the microwave backgrounds \cite{bat, avelino}: the value of $\alpha $ at
the last \ scattering redshift, $z=1000,$ is only 0.005\% lower than its
value today. similarly, the essentially constant evolution of $\alpha $
predicted during the radiation era leads us to expect no measurable effects
on the products of Big Bang nucleosynthesis (BBN) \cite{jdb} because $\alpha 
$ was only 0.007\% smaller at BBN than it is today. This does not rule out
the possibility that unification effects in a more general theory might
require variations in weak and strong couplings, or their contributions to
the neutron-proton mass difference, which might produce observable
differences in the light element productions and new constraints on varying $%
\alpha $ at $z\sim 10^{9}-10^{10}.$ By contrast varying alpha cosmologies
with $\zeta >0$ lead to bad consequences . The fine structure falls rapidly
at late times and the variation is such that it even comes to dominate the
Friedmann equation for the cosmological dynamics. We regard this as a signal
that such models are astrophysically ruled out and perhaps also
mathematically badly behaved.

We should also mention that theories in which $\alpha $ varies will in
general lead to violations of the weak equivalence principle (WEP). This is
because the $\alpha $ variation is carried by a field like $\psi $ and this
couples differently to different nuclei because they contain different
numbers of electrically charged particles (protons). The theory discussed
here has the interesting consequence of leading to a relative acceleration
of order $10^{-13}$ \cite{bmswep} if the free coupling parameter is fixed to
the value given in eq. (\ref{om}) using a best fit of the theories
cosmological model to the QSO observations of refs. \cite{webb1, webb2}.
Other predictions of such WEP violations have also been made in refs. \cite%
{poly, zal, zald, dam}. The observational upper bound on this parameter is
just an order of magnitude larger, at $10^{-12},$ but space-based tests
planned for the STEP mission are expected to achieve a sensitivity of order $%
10^{-18}$ and will provide a completely independent check on theories of
time-varying $e$ and $\alpha .$This is an exciting prospect for the future.

\subsection{The nature of the Friedmann solutions}

The cosmological behaviour of the solutions to these equations was studied
by us in detail, both analytically and numerically in refs. \cite{bsm1,
bsm2, bsm3, bsm4}, \cite{bmota}. Typically, the variation in $\alpha $ does
not have a significant effect on the evolution of the scale factor at late
times although the cosmological expansion does significantly affect the
evolution of $\alpha .$ The evolution of $\alpha $ is summarised as follows:

During the radiation era $a(t)\sim t^{1/2}$ and $\alpha $ is constant in
universes with our entropy per baryon and present value of $\alpha $ like
our own. It increases in the dust era, where $a(t)\sim t^{2/3}$. The
increase in $\alpha $ however, is very slow with a late-time solution for $%
\psi $ proportional to $\frac{1}{2}\log (2N\log (t))$, and so

\begin{equation}
\alpha \sim 2N\log t
\end{equation}

This slow increase continues until the expansion becomes dominated by
negative curvature, $a(t)\sim t$, or by a cosmological vacuum energy, $%
a(t)\sim \exp [\Lambda t/3]$. Thereafter $\alpha $ asymptotes rapidly to a
constant. If we set the cosmological constant equal to zero and $k=0$ then,
during the dust era, $\alpha $ would continue to increase indefinitely. $\ $%
The effect of the expansion is very significant at all times. If we were to
turn it off and set $a(t)$ constant then we could solve the $\psi $ equation
to give the following exponentially growing evolution for $\alpha $, \cite%
{bmota}$:$

\begin{equation}
\alpha =\exp [2\psi ]=A^{-2}\cosh ^{2}[AN^{1/2}(t+t_{0})];\text{ }A\text{
constant.}
\end{equation}

From these results it is evident that non-zero curvature or cosmological
constant brings to an end the increase in the value of $\alpha $ that occurs
during the dust-dominated era. Hence, if the spatial curvature and $\Lambda $
are both too\textit{\ small} it is possible for the fine structure constant
to grow too large for biologically important atoms and nuclei to exist in
the universe. There will be a time in the future when $\alpha $ reaches too
large a value for life to emerge or persist. The closer a universe is to
flatness or the closer $\Lambda $ is to zero so the longer the monotonic
increase in $\alpha $ will continue, and the more likely it becomes that
life will be extinguished. Conversely, a non-zero positive $\Lambda $ or a
non-zero negative curvature will stop the increase of $\alpha $ earlier and
allow life to persist for longer. If life can survive into the curvature or $%
\Lambda $-dominated phases of the universe's history then it will not be
threatened by the steady cosmological increase in $\alpha $ unless the
universe collapses back to high density.

This type of behaviour can also be found in the presence of time-varying $G$%
. If a BD dust universe is exactly flat ($k=0$) then $G$ will continue to
fall forever. Only if there is negative curvature will the evolution of $G$
eventually be turned off and the expansion asymptote to the Milne behaviour
with $a=t$ and $G$ $\rightarrow $ constant. Again, without the small
deviation from flatness the strength of gravity would ultimately become too
weak for the existence of stars and planets and the universe would become
biologically inhospitable, if not uninhabitable.

There have been several studies, following Carter, \cite{car} and Tryon \cite%
{try}, of the need for life-supporting universes to expand close to the
'flat' Einstein de Sitter trajectory for long periods of time. This ensures
that the universe cannot collapse back to high density before galaxies,
stars, and biochemical elements can form by gravitational instability, or
expand too fast for stars and galaxies to form by gravitational instability 
\cite{ch, bt}. Likewise, it was pointed out by Barrow and Tipler, \cite{bt}
that there are similar anthropic restrictions on the magnitude of any
cosmological constant, $\Lambda $. If it is too large in magnitude it will
either precipitate premature collapse back to high density (if $\Lambda <0$)
or prevent the gravitational condensation of any stars and galaxies (if $%
\Lambda >0$). Thus existing studies provide anthropic reasons why we can
expect to live in an old universe that is neither too far from flatness nor
dominated by a much stronger cosmological constant than observed ($%
\left\vert \Lambda \right\vert \leq 10\left\vert \Lambda _{obs}\right\vert $%
).

Inflationary universe models provide a possible theoretical explanation for
proximity to flatness but no explanation for the smallness of the
cosmological constant. Varying speed of light theories \cite{moffat, am, ba,
bm, bhvsl} offer possible explanations for proximity to flatness and
smallness of a classical cosmological constant (but not necessarily for one
induced by vacuum corrections in the early universe). We have shown that if
we enlarge our cosmological theory to accommodate variations in some
traditional constants then\textit{\ it appears to be anthropically
disadvantageous for a universe to lie too close to flatness or for the
cosmological constant to be too close to zero}. This conclusion arises
because of the coupling between time-variations in constants like $\alpha $
and the curvature or $\Lambda $, which control the expansion of the
universe. The onset of a period of $\Lambda $ or curvature domination has
the property of dynamically stabilising the constants, thereby creating
favourable conditions for the emergence of structures. This point has been
missed in previous studies because they have never combined the issues of $%
\Lambda $ and flatness and the issue of the values of constants. By coupling
these two types of anthropic considerations we find that too small a value
of $\Lambda $ or the spatial curvature can be as poisonous for life as too
much. Universes like those described above, with increasing $\alpha (t),$
lead inexorably to an epoch where $\alpha $ is too large for the existence
of atoms, molecules, and stars to be possible.

Surprisingly, there has been almost no consideration of habitability in
cosmologies with time-varying constants since Haldane's discussions \cite%
{hal} of the biological consequences of Milne's bimetric theory of gravity.
Since then, attention has focussed upon the consequences of universes in
which the constants are different but still constant. Those cosmologies with
varying constants that have been studied have not considered the effects of
curvature or $\Lambda $ domination on the variation of constants and have
generally considered power-law variation to hold for all times. The examples
described here show that this restriction has prevented a full appreciation
of the coupling between the expansion dynamics of the universe and the
values of the constants that define the course of local physical processes
within it. Our discussion of a theory with varying $\alpha $ shows for the
first time a possible reason why the 3-curvature of universes and the value
of any cosmological constant may need to be bounded \textit{below} in order
that the universe permit atomic life to exist for a significant period.
Previous anthropic arguments have shown that the spatial curvature of the
universe and the value of the cosmological constant must be bounded \textit{%
above} in order for life-supporting environments (stars) to develop. We note
that the lower bounds discussed here are more fundamental than these upper
bounds because they derive from changes in $\alpha $ which have direct
consequences for biochemistry whereas the upper bounds just constrain the
formation of astrophysical environments by gravitational instability. Taken
together, these arguments suggest that within an ensemble of all possible
worlds where $\alpha $ and $G$ are time variables, there might only be a
finite interval of \textit{non-zero }values of the curvature and
cosmological constant contributions to the dynamics that both allow galaxies
and stars to form and their biochemical products to persist.

\subsection{The role of inhomogeneities}

We can also detect where and how we might expect spatial variations to arise
in a fuller description. Aside from the complexities of the full
inhomogeneous cosmological solution for the formation of galaxies, stars,
and planets, we can isolate non-uniformities that enter through the constant
parameter $N$ which dictates the form and time-evolution of $a(t)$ and $%
\alpha (t).$ First we see that $N$ is proportional to the density of
electromagnetically charged matter in the universe. This will possess some
spatial variation and is of order $10^{-5}$ on large scales. More
significant though is the variation of the baryonic content of the CDM
density with scale. We need the CDM to be dominated by matter with magnetic
charge (but see Bekenstein). This can be the case on large scale but we know
that the dark matter becomes dominated by baryons (therefore with $\zeta >0$%
) locally. Hence, there is expected to be a very significant spatial
variation of $\zeta $ with scale, including a change of sign, which will
feed into the variation of $\alpha .$

\subsection{\protect\bigskip General properties of the evolution of alpha
and G}

The evolution equation for $\psi (t)$ has a number of simple but important
properties. Since $N>0$ the right-hand side or eq. (\ref{psidot2}) must be
positive. This means that linearisations of this equation are dangerous and
give rise to linearisation instabilities unless attention is confined to the
regime $\psi <<1.$ In general the positivity property means that there can
be no oscillations of $\psi $ or $\alpha $ in time in solutions of this
equation. This follows from the required positivity of \ $(\dot{\psi}a^{3}%
\dot{)}$, which means that $\psi $cannot have a maximum. The possible
cosmological evolutions for $\psi $ and $\alpha $ are decrease to a minimum
followed by a monotonic increase, monotonic decrease, or monotonic increase.
This conclusion holds independently of the value of $k$ in the Friedmann
equation. This has one very important consequence. It means that the
asymptotic monotonic non-decrease of $\alpha $ found in our flat and open
universes will still occur in closed universes. There cannot be a sudden
change in the evolution of $\alpha $ when the universe starts to collapse.
This also means that if we model spherical overdensities by closed universes
embedded in a flat background then the evolution of $\alpha (t)$ in the
overdensities will be very similar to that in the flat background even when
the overdensities collapse to form bound 'clusters'. This has the important
implication that such an inhomogeneous universe will not end up with very
different values of $\alpha $ and $\dot{\alpha}$ in inside and outside the
bound inhomogeneities.

This argument can also be applied to the evolution of $\phi $ and $G$ in BD
theory. Consider the case of dust ($p=0$). The combination $(\dot{\phi}a^{3}%
\dot{)}$ must now be positive and so $\phi $ cannot have a maximum and $G$
cannot have a minimum regardless of the sign of the curvature parameter $k$.
In particular, $G(t)$ cannot oscillate. Again, this property acts as a
safeguard on the divergent evolution of $G$ inside and outside overdensities.

\section{The Second Law}

There has been considerable recent discussion \cite{dav, duff} about the
equivalence of models of the variation of different dimensional 'constants'
of Nature. In particular, it has been suggested that consideration of the
second law of black hole thermodynamics distinguishes, say, variations of $e$
from variations of $c$ and that some of these variations could be ruled out
because they bring about a decrease in time of the Bekenstein-Hawking
entropy of a charged black hole. Others have argued that no such distinction
is operationally possible. However, we believe that the most crucial factor
has been missed in this discussion. In theories which generalise general
relativity by allowing traditional constants (like $G$ or $e$) to vary the
black hole solutions with event horizons are particular solutions of the
theory in which the constant concerned is a constant. When the constant
varies the black hole solution no longer exists and there is no longer any
black hole thermodynamics to constrain the variation. The situation is very
clear in the simple case of a Schwarzschild black hole in Brans Dicke
theory. We know from the work of Hawking \cite{swh} that the black hole
solutions are the same as those in general relativity. Thus Schwarzschild is
a $\phi \symbol{126}G^{-1}=$ constant solution of the Brans-Dicke field
equations. The entropy of this black hole is

\begin{equation*}
S_{bh}\symbol{126}GM^{2}.
\end{equation*}%
If we were to apply the second law to this formula it would appear to say
that all cosmological solutions in which $G$ falls with time are ruled out.
However, this would not be a correct deduction (which is fortunate because
we see from eq. (\ref{bds2}) that essentially all Brans-Dicke cosmologies
have such behaviour) because $\phi $ and $G$ are\textit{\ constant} on the
Schwarzschild horizon. If we allow variation of $G$ then the solution turns
into a naked singularity and the thermodynamic relations no longer exist.
Thus one cannot at present use considerations of black hole thermodynamics
to constrain or distinguish the time or space variation of constants of
Nature by simply 'writing in' time variations into the formulae that define
the black hole when these constants do not vary.

\textbf{Acknowledgements} I would like to thank my collaborators Jo\~{a}o
Magueijo, H\aa vard Sandvik, John Webb, Michael Murphy and David Mota for
their essential contributions to the work described here. I would also like
to thanks Bruce Bassett, Thibault Damour, Paul Davies, Tamara Davies, Carlos
Martins, John Moffat and Clifford Will for discussions.

\end{document}